\documentclass[12pt]{article}

\usepackage{newtxtext,newtxmath}
\usepackage[utf8x]{inputenc}
\usepackage[english]{babel}
\usepackage{amsmath,amsbsy,amsfonts}
\usepackage{graphicx}
\usepackage{subfig}
\usepackage{color}
\usepackage{float}
\usepackage{cite}
\usepackage[left=1cm,right=1cm,top=2cm,bottom=2cm]{geometry}
\usepackage{physics}
\usepackage[colorlinks=true, allcolors=red]{hyperref}

\numberwithin{equation}{section} 

\begin{document}

\title{Approximate solutions for the ion-laser interaction in the high intensity regime: Matrix method perturbative analysis}

\author{B.M. Villegas-Martínez$^*$, H.M. Moya-Cessa and F. Soto-Eguibar\\
	\small Instituto Nacional de Astrofísica, Óptica y Electrónica, INAOE \\
	\small {Calle Luis Enrique Erro 1, Santa María Tonantzintla, Puebla, 72840 Mexico}\\
	$^*$\small {Corresponding author: bvillegas@inaoep.mx}}

\date{\today}

\maketitle

\begin{abstract}
We provide an explicit expression for the second-order perturbative solution of a single trapped-ion interacting with a laser field in the strong excitation regime. From the perturbative analytical solution, based on a matrix method and a final normalization of the perturbed solutions, we show that the probability to find the ion in its excited state fits well with former results.
\end{abstract}

\section{Introduction}
Trapped ions interacting with laser beams represent a quantum optical elementary system that has gained considerable interest in quantum information, both experimentally and theoretically, for its potential to realize quantum computation. Their undoubted popularity relies on the preparation of nonclassical states of the ion's vibrational motion \cite{1A,2A,3A,4A,5A,6A,A7.1,A7.2},  the reconstruction of quasiprobability distribution functions \cite{A7}, the production of quantum logic gates and the preparation of entangled Bell states \cite{A8,A9,A10} in quantum computers, and so forth.\\
In spite that a single trapped ion interacting with one laser beam constitutes a simple model, the theoretical treatment of its dynamics is a non-trivial task and its study involves the adoption of suitable approximations, for instance, the optical rotating wave approximation (RWA), the Lamb-Dicke approximation\cite{A11}, in which the ion moves within a region much smaller than the laser wavelength, the vibrational RWA  \cite{A12}, where the counter-rotating terms are neglected and the weak excitation regime\cite{A13,A14}, in which the amplitude of the laser intensity is much smaller than the vibrational frequency of the ion. The latter one has a severe limit because it ignores the case for an intense laser field that becomes crucial due its potential applications for faster quantum gates \cite{A15,A16,A17}. \\
Lately, several perturbation techniques have been applied to find solutions in the high intensity regime \cite{A18,A19,A20}; however, their treatment resides in a somewhat complicated unitary set of transformations to obtain effective Hamiltonians that have a diagonal form \cite{A18,A19} or a large perturbative decomposition of the evolution operator in a generalized Magnus expansion\cite{A20}. This prompt us to explore a straightforward alternative to look for perturbative solutions of the Schrödinger equation for a single trapped ion, valid for large  laser intensities. Hence, in this work we apply a simple perturbative approach, namely the Normalized Matrix Perturbation Method (NMPM) \cite{A21,A22,A23,A24,A25}, and which unlike all other standard perturbation approaches,  has a great freedom to choose what part to the ion trap Hamiltonian represents the perturbation, a duality that allows us the possibility to analyze the system in the high intensity regime. Additionally, the NMPM is capable of providing normalized solutions at any correction order, a property that marks the difference between our approach and those already existing in the literature, where the general formulation of a normalization constant isnot easy to study \cite{A26,A27,A28,A29,A30,A31}. Because the low intensity has been extensively studied in the literature \cite{1A,2A,3A,4A,5A,6A,A7.1} by making use of the RWA, in this contribution we solve, by using the NMPM, the high intensity regime up to second order. As the ion-laser interaction Hamiltonian has a similarity transformation with the Rabi Hamiltonian \cite{A7.2}, we also outline  the second order solution of that problem.

\section{Outline of the method}
The NMPM is a time-independent perturbative approach \cite{A21,A22,A23,A24,A25} explicitly based on a Taylor expansion of the evolution operator of the formal solution of the time-dependent Schrödinger equation $\left|\psi(t)\right \rangle=e^{-i \hat{H} t} \left|\psi(0) \right \rangle$, whose Hamiltonian $\hat{H}$ is the sum of an already solved part $\hat{H}_0$ plus a perturbation $\lambda\hat{H}_p$, where $\lambda$ is a real and dimensionless perturbation parameter that establishes the order of the perturbation. For example, an expansion for the propagator truncated up to first-order is 
\begin{equation} \label{a1}
\left|\psi(t) \right \rangle=  \left[ e^{-i \hat H_0 t} + \lambda \sum\limits_{n=1}^{\infty} \frac{(-it)^n}{n!}  \sum\limits_{k=0}^{n-1} \hat{H}_{0}^{n-1-k} \hat{H_p} \hat{H}_{0}^{k}\right]     \left| \psi(0)\right \rangle.
\end{equation}
The main ingredient of the approach NMPM \cite{A21,A22,A24} is centred on the implementation  of the upper triangular matrix
\begin{equation} \label{a2}
M=
\begin{pmatrix}
\hat{H}_{0} & \hat{H_p}  \\
0 & \hat{H}_{0} \\
\end{pmatrix},
\end{equation}
whose diagonal elements are conformed by the unperturbed Hamiltonian and the upper triangle by the perturbation. If we multiply the matrix $M$ by itself $n$-times, we found that its upper element contains exactly the same products of $\hat H_0$ and $\hat{H_p}$ as the summation in equation \eqref{a1}. In short, the matrix element $M_{1,2}$  gives us the first order correction; based on this consideration, equation \eqref{a1} may be  rewritten as
\begin{equation} \label{a3}
\left|\psi(t) \right \rangle= \left[ e^{-i \hat{H}_{0}} + \lambda (e^{-i M t})_{1,2}\right]  \left| \psi(0)\right \rangle.
\end{equation}
Therefore, the approximate solution has been split in two parts; the first one being the solution of the unperturbed system, that is the one we know, while the second one refers to the first-order correction. In order to determine the solution to first order, we have to keep in mind that the problem originally posed must follows the same matrix convention; hence, the approximated solution \eqref{a3} may be conveniently written as
\begin{equation} \label{a4}
\left|\psi(t) \right \rangle=  \left| \psi^{(0)} \right \rangle + \lambda \left( \left| \psi^{(1)} \right \rangle\right)_{1,2},
\end{equation}
where the superscripts denote the order of the correction and  $\left| \psi^{(1)} \right \rangle$ is the matrix defined by
\begin{equation} \label{a5}
\left| \psi^{(1)} \right \rangle=
\begin{pmatrix}
\left| \psi_{1,1} \right \rangle & \left| \psi_{1,2} \right \rangle  \\
\left| \psi_{2,1} \right \rangle & \left| \psi_{2,2} \right \rangle \\
\end{pmatrix}.
\end{equation}
The solution to first order may be determined by deriving the equations \eqref{a3} and \eqref{a4} with respect to time, and equating the corresponding powers of $\lambda$ and performing the algebraic steps outlined in \cite{A21,A22,A24}. Then we obtain
\begin{equation} \label{a6}
\ket{\psi_{1,2}} = -i e^{-i \hat{H}_{0} t}
\left[\int\limits_0^t  e^{i \hat{H}_{0} t_1} \hat{H}_p e^{-i \hat{H}_{0}t_1}  dt_1 \right] \ket{\psi(0)}.
\end{equation}
All the information of the second-order correction will be in the element $M_{1,3}$ of a newly defined 3×3 triangular matrix $M$, completely similar to \eqref{a2}. Thus, it becomes clear that the matrix treatment allows to transform the Taylor series of the formal solution of the time-dependent Schrödinger equation in a power series of the matrix $M$, where the kets $\left| \psi^{(k)} \right \rangle$ are obtained iteratively and may be easily handled. Therefore, the corresponding relation that allows us to find perturbative solutions in the Schrödinger equation at $k$-th correction order  is given by \cite{A21,A22,A24}
\begin{align}\label{a7}
\left|\Psi(t) \right \rangle =& N^{(k)}(t) \left[ \left| \psi^{(0)} \right \rangle + \sum\limits_{n = 1}^{k} \lambda^n \left( \left| \psi^{(n)} \right \rangle\right)_{1,n+1} \right],
\end{align}
where $N^{(k)}(t)$ is a normalization factor that preserves the norm at any order and that is defined as
\begin{align}\label{a8} 
N^{(k)}(t)^{-2} =& 1 + 2 \sum\limits_{n = 1}^{k} \lambda^n \; \Re \left( \left\langle \psi^{(0)} |  \psi_{1,n+1} \right \rangle \right)   + \sum\limits_{n = 1}^{k}\lambda^{2n} \left\langle \psi_{1,n+1} | \psi_{1,n+1} \right\rangle \nonumber\\
& +2 \sum_{\substack{n=1\\ k>1}}^{k-1} \sum\limits_{m = 2n+1}^{n+k} \lambda^{m} \; \Re \left( \left\langle \psi_{1,n+1} | \psi_{1,m-n+1} \right\rangle \right), 
\end{align}
where $\Re\left(z\right)$ means the real part of $z$. The matrix element $\left|\psi_{1,n+1} \right \rangle$ is the relevant solution we are looking for and is expressed in the form
\begin{align}\label{a9}
\left|\psi_{1,n+1} \right \rangle =& \left(-i\right)^n e^{-i \hat H_0 t}\int\limits_0^t  dt_1 \int\limits_0^{t_1 }  dt_2 \dots \int\limits_0^{t_{n-1} }  dt_n \ e^{i \hat H_0 t_1}  \hat{H_p} e^{-i \hat H_0 t_1}  e^{i \hat H_0 t_2}\hat{H_p} e^{-i \hat H_0 t_2}  \dots e^{i \hat H_0 t_n}  \hat{H_p} e^{-i \hat H_0 t_n}   \left|\psi(0) \right \rangle.
\end{align}
The normalized solution \eqref{a7}, that contains $N^{(k)}(t)$, has been obtained and published separately; the reader can consult reference \cite{A24} for further details. From here, expressions \eqref{a7}, \eqref{a8} and \eqref{a9} are the equations that we will use to obtain the normalized perturbative solutions for the trapped ion-laser system in the high intensity regime.

\section{Trapped-ion Hamiltonian}
We now consider a simplified model of a single trapped ion interacting with a classical laser field that is described by the  Hamiltonian \cite{A32}
\begin{equation} \label{b1}
\hat{H}_{\textrm{ion}}=\nu \hat{n}  + \frac{\delta}{2}\hat{\sigma_{z}} +\Omega \left[ \hat{\sigma}^{+} \hat{D}(i \eta) + \hat{\sigma}^{-} \hat{D}^\dagger (i \eta)\right],
\end{equation}
where $\hat{D}\left(i\eta\right)=\exp[i\eta \left(\hat{a} + \hat{a}^{\dagger}\right)]$ is the Glauber displacement operator, being $\hat{a}^{\dagger}$ ($\hat{a}$) the ion’s vibrational creation (annihilation) operator with $\hat{n}=\hat{a}^{\dagger}\hat{a}$ the number operator, $\eta$ is the so-called Lamb-Dicke parameter, $\nu$ is the trapping frequency, $\delta=\nu \kappa$ is the laser-ion detuning  which is considering here as a multiple integer of the vibrational frequency of the ion, $\Omega$ is the Rabi frequency, i.e. the ion-laser coupling strength, and $\hat{\sigma}^{+}=\ket{\uparrow}\bra{\downarrow}$ and $\hat{\sigma}^{-}=\ket{\downarrow}\bra{\uparrow}$ are the atomic raising and lowering operators (Pauli matrices) expressed in terms of the excited $\ket{\uparrow}=(1,0)$ and ground $\ket{\downarrow}=(0,1)$ states of the two-level ion, which obey the commutation relations  $\left[\hat{\sigma}^{+}, \hat{\sigma}^{-}\right]=\hat{\sigma_{z}}$ and $\left[\hat{\sigma_{z}}, \hat{\sigma}^{\pm} \right]=\pm 2 \hat{\sigma}^{\pm}$, respectively. The dynamics of a single trapped ion can be studied by solving the time-dependent Schrödinger equation 
\begin{equation} \label{b2}
i\frac{d}{dt}\ket{\psi(t)}=\hat{H}_{\textrm{ion}}\ket{\psi(t)}.
\end{equation}
We are interested in solving perturbatively the Schrödinger equation through the NMPM. Indeed, our perturbative scheme provides us the great flexibility to choose the unperturbed and perturbed parts of the full Halmiltonian \eqref{b1}. Using this freedom, we can get normalized perturbative solutions for the cases when the amplitude $\Omega$ of the laser is very small compared with the vibrational frequency $\Omega\ll\nu$ of the ion and vice versa, i.e. when $\Omega\gg\nu$; these two approximations correspond to the weak and strong laser intensity regimens. In addition, the ion-trap system is formally equivalent to the quantum Rabi model when we consider a certain unitary transformation $\hat{T}$; therefore, we can perform the transformation $\ket{\phi(t)}=\hat{T}\ket{\psi(t)}$ and also get perturbative solutions of the quantum Rabi model for the weak $\left(g/\omega\right)\ll 1$ and strong $\left(g/\omega\right)\gg1$ coupling regime; this will be done below in Section 5.

\section{High intensity regime}
\subsection{First order correction}
Let us begin our perturbative analysis by solving the high intensity case $(\Omega/\nu\gg 1)$. In such scenario, we must consider that $\hat{H}_{p}=\hat{n} + \frac{\kappa}{2} \hat{\sigma_{z}}$ is the perturbation with perturbative parameter $\lambda=\nu/ \Omega$, whereas $\hat{H}_{0}= \hat{\sigma}^{+} \hat{D}(i \eta) + \hat{\sigma}^{-} \hat{D}^\dagger (i \eta)$ plays the role of the unperturbed part. If we re-scale time as $\tau=\Omega t$ and setting $k=1$ into Eq.~\eqref{a7}, we arrive to the set of equations to get the approximate solution at first order
\begin{equation} \label{c1}
\ket{\psi(\tau)}\approx N^{(1)}(\tau)\left[\ket{\psi^{(0)}}+ \lambda \ket{\psi^{(1)}} \right],
\end{equation}
where 
\begin{subequations}
\begin{align} \label{c2a}
\ket{\psi^{(0)}}&= e^{-i\hat{H}_{0}\tau} \ket{\psi(0)}, \\ \label{c2b}
\ket{\psi^{(1)}}&= -i e^{-i\hat{H}_{0}\tau} \int \limits_{0}^{\tau} e^{i \hat{H}_{0} \tau_{1}} \left(\hat{n} + \frac{\kappa}{2} \hat{\sigma_{z}}\right) e^{-i \hat{H}_{0} \tau_{1}} d\tau_{1} \ket{\psi(0)}, \\ \label{c2c}
 N^{(1)}(\tau)^{-2}&=1+ 2 \lambda \Re \left(\braket{\psi^{(0)}}{\psi^{(1)}}\right)  +\lambda^2 \braket{\psi^{(1)}}{\psi^{(1)}}.
\end{align}
\end{subequations}
The integral in Eq.~\eqref{c2b} requires to calculate the product of exponential operators with $\hat{n} + \frac{\kappa}{2} \hat{\sigma_{z}}$. To do this, we first expand in Taylor series the exponential operator $e^{i \hat{H}_{0} \tau_{1}}$ and split the series in even and odd powers of $\hat{H}_{0}$,
\begin{align} \label{c3}
e^{i \hat{H}_{0} \tau_{1}}&= \displaystyle\sum_{n=0}^\infty \frac{(-1)^n \tau^{2 n}_{1}}{(2 n)!} \hat{H}^{2n}_{0} + i \displaystyle\sum_{n=0}^\infty \frac{(-1)^n \tau^{2 n+1}_{1}}{(2 n+1)!} \hat{H}^{2n+1}_{0};
\end{align}
one can easily check that $\hat{H}^{2 n}_{0}=\hat{1}$ and $ \hat{H}^{2 n +1 }_{0}= \hat{H}_{0}$, then equation Eq.~\eqref{c3} becomes
\begin{align} \label{c4}
e^{i \hat{H}_{0} \tau_{1}}=& \cos(\tau) + i \sin(\tau) \left[\hat{\sigma}^{+} \hat{D}(i\eta) + \hat{\sigma}^{-} \hat{D}^\dagger (i \eta) \right].
\end{align}
It is possible to show that
\begin{align}  \label{c5}
e^{i \hat{H}_{0} \tau_{1}} \left(\hat{n} + \frac{\kappa}{2} \hat{\sigma_{z}}\right) e^{-i \hat{H}_{0} \tau_{1}}&= \hat{n} \cos^2(\tau_{1})+ \frac{i}{2} \sin(2\tau_{1})  \left[\hat{H}_{0}, \hat{n} \right] + \sin^2 \left( \tau_{1} \right) \hat{H}_{0} \hat{n} \hat{H}_{0} + \frac{\kappa}{2} \cos(2\tau_{1}) \hat{\sigma_{z}} + i\frac{\kappa}{2} \sin(2\tau_{1}) \hat{H}_{0} \hat{\sigma_{z}}.
\end{align}
As $\left[\hat{a}^{\dagger},\hat{n}\right]=-\hat{a}^{\dagger}$ and , $\left[\hat{a},\hat{n}\right]=\hat{a}$, we obtain that
\begin{align}  \label{c6}
\left[\hat{H}_{0}, \hat{n} \right]=& \hat{\sigma}^{+} \left[ \hat{D}(i \eta ), \hat{n} \right] + \hat{\sigma}^{-} \left[ \hat{D}^\dagger(i \eta ), \hat{n} \right] \nonumber\\
=& \hat{\sigma}^{+} \hat{D}(i \eta )\left[ \hat{n}-\hat{D}^\dagger(i \eta ) \hat{n}  \hat{D}(i \eta ) \right] + \hat{\sigma}^{-} \hat{D}^\dagger(i \eta )\left[ \hat{n}-\hat{D}(i \eta )\hat{n} \hat{D}^\dagger(i \eta )\right].
\end{align}
Using the Hadamard formula \cite{A33}, $e^{\delta \hat{A}} \hat{B} e^{-\delta \hat{A}} =\hat{A} + \delta \left[\hat{A}, \hat{B}\right] + \frac{\delta^2}{2!} \left[\hat{A},\left[\hat{A},\hat{B}\right]\right]+\hdots$ , expression \eqref{c6} is simplified to
\begin{align}  \label{c7}
\left[\hat{H}_{0}, \hat{n} \right]
=& -\eta \Big[\hat{\sigma}^{+} \hat{D}(i\eta) + \hat{\sigma}^{-} \hat{D}^\dagger (i \eta) \Big] \left[\eta+ i \left(\hat{a}-\hat{a}^\dagger \right) \hat{\sigma_{z}}  \right].
\end{align}
At this point, one can prove that
\begin{align} \label{c8}
\hat{H}_{0} \hat{n}\hat{H}_{0}=& \begin{pmatrix}
\hat{D}(i \eta ) \hat{n} \hat{D}(i \eta )^\dagger & 0 \\ 0 & \hat{D}(i \eta )^\dagger \hat{n} \hat{D}(i \eta ) 
\end{pmatrix} = \hat{n} + \eta^2 + i \eta \left(\hat{a}-\hat{a}^\dagger\right) \hat{\sigma}_{z},
\end{align}
and when this result is substituted in Eq.~\eqref{c5} together with Eq~\eqref{c8} yields to
\begin{align}  \label{c9}
e^{i \hat{H}_{0} \tau_{1}} \left(\hat{n} + \frac{\kappa}{2} \hat{\sigma_{z}}\right) e^{-i \hat{H}_{0} \tau_{1}}  =&\hat{n}  -\frac{i \eta}{2 } \Big[\hat{\sigma}^{+} \hat{D}(i\eta) + \hat{\sigma}^{-} \hat{D}^\dagger (i \eta) \Big] \left[\eta+ i \left(\hat{a}-\hat{a}^\dagger \right) \hat{\sigma_{z}} -\frac{\kappa}{\eta} \hat{\sigma_{z}} \right]  \sin(2\tau_{1}) + \frac{\kappa}{2} \cos\left(2\tau_{1} \right) \hat{\sigma_{z}} \nonumber\\
& + \eta \left[\eta + i\left(\hat{a}-\hat{a}^{\dagger}\right) \hat{\sigma}_{z}\right] \sin^2 \left(\tau_{1} \right),
\end{align}
that can be easily integrated to give
\begin{align} \label{c10}
\int \limits_{0}^{\tau} e^{i \hat{H}_{0} \tau_{1}} \left(\hat{n} + \frac{\kappa}{2} \hat{\sigma_{z}}\right) e^{-i \hat{H}_{0} t_{1}} \tau_{1}=& \bigg\lbrace 2 \hat{n} + \eta\left[\eta + i \left(\hat{a}-\hat{a}^{\dagger} \right) \hat{\sigma}_{z}\right] \bigg\rbrace  \frac{\tau}{2} - \frac{\eta}{4} \left[\eta + i\left(\hat{a}-\hat{a}^{\dagger}\right) \hat{\sigma}_{z}  -\frac{\kappa}{\eta} \hat{\sigma}_{z} \right] \sin \left(2 \tau \right)   \nonumber\\
& -\frac{i \eta}{2 } \Big[\hat{\sigma}^{+} \hat{D}(i\eta) + \hat{\sigma}^{-} \hat{D}^\dagger (i \eta) \Big] \left[\eta+ i \left(\hat{a}-\hat{a}^\dagger \right) \hat{\sigma_{z}} -\frac{\kappa}{\eta} \hat{\sigma_{z}} \right]  \sin^2(\tau),
\end{align}
and by substituting into Eq.~\eqref{c2b} and after some algebra, the first order term is obtained
\begin{align} \label{c11}
\left|\psi^{(1)} \right \rangle=& -i\cos(\tau) \bigg\lbrace \hat{n} \tau + \frac{\eta}{2} \left[\tau-\tan(\tau)\right] \left[\eta + i  \left(\hat{a}-\hat{a}^{\dagger}\right) \hat{\sigma}_{z} \right] + \frac{\kappa}{2} \tan\left(\tau\right) \hat{\sigma_{z}} \bigg\rbrace \left|\psi(0) \right \rangle \nonumber\\
&-\tau \sin(\tau) \left[\hat{\sigma}^{+} \hat{D}\left(i\eta\right)  +\hat{\sigma}^{-} \hat{D}^{\dagger}\left(i\eta\right) \right] \bigg\lbrace \hat{n} + \frac{\eta}{2} \left[\eta + i \left(\hat{a}-\hat{a}^{\dagger}\right) \hat{\sigma_{z}}\right]  \bigg\rbrace \left|\psi(0) \right \rangle.
\end{align}
Now, the normalization constant $N^{(1)}(\tau)$ of Eq.~\eqref{c2c} is obtained doing the inner product of $\left|\psi^{(1)} \right \rangle$ with itself, and once it calculated and after being substituted in Eq.~\eqref{c1} give us 
\begin{align} \label{c12}
\ket{\psi(\tau)}  \approx & N^{(1)}(\tau) \cos(\tau) \bigg\lbrace 1-i\lambda \tau \hat{n} -i \lambda \frac{\eta }{2} \left[ \tau-\tan(\tau)\right] \left[\eta + i  \left(\hat{a}-\hat{a}^{\dagger}\right) \hat{\sigma}_{z} \right] -i\frac{\lambda \kappa}{2} \tan\left(\tau\right) \hat{\sigma_{z}} \bigg\rbrace \ket{\psi(0)} \nonumber\\
&-i N^{(1)}(\tau) \sin(\tau) \left[\hat{\sigma}^{+} \hat{D}\left(i\eta\right) +\hat{\sigma}^{-} \hat{D}^{\dagger}\left(i\eta\right) \right] \left[1 + \lambda \frac{\eta \tau}{2} \left(\hat{a}-\hat{a}^{\dagger}\right) \hat{\sigma_{z}}-i\lambda \tau\left(\hat{n}+\eta^2/2\right)\right]  \ket{\psi(0)}.
\end{align}
The above expression is the first order approximated solution of Eq.~\eqref{b2} and the normalization constant is
\begin{align}  \label{c13}
\left[N^{(1)}(\tau)\right]^{-2}=&1+ \frac{\lambda^2 \eta^2 }{4} \left(\eta^2+1\right) \Bigg\lbrace \sin^2(\tau) + \tau \left[\tau-\sin(2\tau)\right]  \Bigg\rbrace +\frac{\lambda^2\kappa^2}{4}\sin^2\left(\tau\right)+ \lambda^2 \tau^2 \bra{\psi(0)}\hat{n}^2 \ket{\psi(0)}  \nonumber\\
& + \frac{\lambda^2 \eta^2}{4} \Bigg\lbrace \sin^2(\tau) + \tau \left[\tau-\sin(2\tau)\right]  \Bigg\rbrace  \bra{\psi(0)} \Bigg[ 2\hat{n}-\left(\hat{a}^2+\hat{a}^{\dagger2}\right) +2i\eta \left(\hat{a}-\hat{a}^{\dagger}\right)\hat{\sigma_{z}}\Bigg]  \ket{\psi(0)} \nonumber\\
&  + \frac{\lambda^2 \tau \eta}{4} \left[2\tau-\sin(2\tau)\right] \bra{\psi(0)} \Bigg\lbrace 2\eta  \hat{n}+ i  \left[2\left( \hat{a}-\hat{a}^{\dagger} \right) \hat{n} -\left(\hat{a}+\hat{a}^{\dagger}\right)\right] \hat{\sigma_{z}} \Bigg\rbrace \ket{\psi(0)} \nonumber\\
& + \frac{\lambda^2 \kappa}{2} \sin\left(2\tau\right) \bra{\psi(0)}  \Big\lbrace \hat{n}\tau + \frac{\eta}{2}\left[\tau-\tan(\tau)\right] \left[\eta + i \left(\hat{a}-\hat{a}^{\dagger}\right)\hat{\sigma_{z}}\right] \Big\rbrace \hat{\sigma_{z}} \ket{\psi(0)}.
\end{align}
Let us consider as initial state $\left|\psi(0) \right \rangle=\ket{n}\ket{g}$, which represents the $n$ vibrational quanta and the ion in the ground internal state $\ket{g}$. With this initial state the solution to first-order is
\begin{align}\label{c14}
\ket{\psi(\tau)}_{n,g}\approx & N^{(1)}_{n,g}(\tau) \cos(\tau) \Bigg\lbrace 1 -i\lambda \tau n-i \lambda \frac{\eta^2}{2} \left[\tau-\tan(\tau)\right] + i\frac{\lambda \kappa}{2} \tan(\tau) \Bigg\rbrace \ket{n}\ket{g}  \nonumber \\
& -\frac{\lambda \eta}{2}  N^{(1)}_{n,g}(\tau) \cos(\tau) \left[\tau-\tan(\tau)\right] \left[\sqrt{n} \ket{n-1} -\sqrt{n+1} \ket{n+1}\right]\ket{g} \nonumber\\
&-i N^{(1)}_{n,g}(\tau) \sin(\tau) \left[1-i\lambda \tau \left(n+\eta^2/2\right)\right] \ket{i\eta;n}\ket{e} \nonumber\\
& + i\frac{ \lambda \eta \tau}{2}   N^{(1)}_{n,g}(\tau) \sin(\tau) \left[\sqrt{n} \ket{i\eta; n-1}-\sqrt{n+1} \ket{i\eta; n+1}\right] \ket{e},
\end{align}
where $\ket{\alpha; k}\equiv \hat{D}\left(\alpha\right)\ket{k}$ is a displaced number state \cite{A33}, whereas the normalization constant $ N^{(1)}_{n,g} (\tau)$ is given by
\begin{align} \label{c15}
\left[ N^{(1)}_{n,g}(\tau) \right]^{-2}
=& 1+\frac{\lambda^2}{8} \bigg\lbrace 4 \tau^2 n \left(\eta^2+2n\right) + 2 \left[ \eta^2 \left(\eta^2+2n+1\right) + \kappa^2\right] \sin^2(\tau) + 2 \tau \eta^2 \left(\eta^2 +4n+1\right) \left[\tau-\sin(2\tau)\right] \bigg\rbrace \nonumber\\
& - \frac{\lambda^2 \kappa}{4} \sin(2\tau) \Big\lbrace  2 n \tau + \eta^2 \left[\tau-\tan(\tau) \right] \Big\rbrace.
\end{align}
If we suppose an initial condition with the ion in the excited state, $\ket{\psi(0)}=\ket{n}\ket{e}$, we get
\begin{align} \label{c16}
\ket{\psi(\tau)}_{n,e}\approx & N^{(1)}_{n,e}(\tau) \cos(\tau) \Bigg\lbrace 1 -i \lambda \tau n-i \lambda \frac{\eta^2}{2} \left[\tau-\tan(\tau)\right] - i\frac{\lambda \kappa}{2} \tan(\tau) \Bigg\rbrace \ket{n}\ket{e} \nonumber \\
& +\frac{\lambda \eta}{2}  N^{(1)}_{n,e}(\tau) \cos(\tau) \left[ \tau-\tan(\tau)\right] \left[\sqrt{n} \ket{n-1} -\sqrt{n+1} \ket{n+1}\right]\ket{e} \nonumber\\
&-i  N^{(1)}_{n,e}(\tau) \sin(\tau) \left[1-i \lambda \tau \left(n+\eta^2/2\right)\right] \ket{-i\eta;n}\ket{g} \nonumber\\
& - i\frac{ \lambda \tau \eta}{2}   N^{(1)}_{n,e}(\tau) \sin(\tau) \left[\sqrt{n} \ket{-i\eta; n-1}-\sqrt{n+1} \ket{-i\eta; n+1}\right] \ket{g} 
\end{align}
where normalization constant is $\left[N^{(1)}_{n,e}(\tau)\right]^{-2}= \left[N^{(1)}_{n,g}(\tau)\right]^{-2} + \frac{\lambda^2 \kappa}{2} \sin(2\tau) \Big\lbrace  2 n \tau + \eta^2 \left[\tau-\tan(\tau) \right] \Big\rbrace$.\\ 
As a third initial condition, we assume $\ket{\psi(0)}=\ket{i\alpha}\ket{e}$ which mean an initial vibrational coherent state and where the ion is initially in its excited state, then Eq.~\eqref{c6} becomes
\begin{align}\label{c17}
\ket{\psi(\tau)}_{i\alpha, e}=& N^{(1)}_{i\alpha,e} \cos(\tau) \bigg\lbrace 1+  i\frac{\lambda \eta}{2} \left(\alpha-\eta\right)  \left[\tau-\tan(\tau)\right]  - i\frac{\lambda \kappa}{2} \tan(\tau)  \bigg\rbrace \ket{i\alpha} \ket{e} \nonumber\\
& -i\frac{\lambda}{2} N^{(1)}_{i\alpha,e} (\tau) \cos(\tau) \bigg\lbrace 2\alpha \tau-\eta \left[\tau-\tan(\tau)\right] \bigg\rbrace \left(\frac{\partial}{\partial \alpha} + \alpha\right) \ket{i\alpha}\ket{e} \nonumber\\
& -\frac{\lambda \tau}{2} N^{(1)}_{i\alpha,e}(\tau) \left(2\alpha-\eta\right) \sin(\tau)  \left[\frac{\partial}{\partial \left(\alpha-\eta\right)} + \alpha-\eta\right] \ket{i\left(\alpha-\eta\right)} \ket{g} \nonumber\\
& - i N^{(1)}_{i\alpha,e} (\tau) \sin(\tau)\left[1-i \frac{\alpha \eta \lambda \tau}{2}\right]  \ket{i\left(\alpha-\eta\right)} \ket{g},
\end{align}
with
\begin{align} \label{c18}
\left[ N^{(1)}_{i\alpha, e} (\tau)\right]^{-2}
 =& 1 + \frac{\lambda^2}{2} \bigg \lbrace  \alpha \eta \left(\alpha-\eta\right) \left(\eta-4\alpha \tau^2\right) + 2 \alpha \tau^2 \left[\alpha\left(\alpha^2 +1\right)-\eta\left(\eta^2+1\right)\right]\bigg\rbrace + \frac{\lambda^2 \kappa^2}{4} \sin^2(\tau)\nonumber\\
& -\frac{\lambda^2}{8}  \bigg \lbrace \eta^2 \left[\left(2\alpha-\eta\right)^2 +1\right] \cos(2\tau)-2 \tau \eta \left(2\alpha-\eta\right)\left[\eta^2+2\alpha\left(\alpha-\eta\right)+1\right]\sin(2\tau) \bigg\rbrace \nonumber\\
& +\frac{\lambda^2 \kappa}{4} \sin(2\tau) \Big\lbrace 2\alpha^2 \tau-\left(2\alpha-\eta\right) \eta  \left[\tau-\tan(\tau)\right]  \Big\rbrace  + \frac{\lambda^2 \eta^2}{8} \left(\eta^2+1\right)\left(2 \tau^2 +1\right),
\end{align}
where we have used the coherent states 
properties $\hat{a} \ket{\beta}=\beta \ket{\beta}$, $\hat{a}^{\dagger}\ket{\beta}=\left(\frac{\partial}{\partial \beta} + \frac{\beta^{*}}{\beta} \frac{\partial}{\partial \beta^{*}}  + \beta^{*}\right) \ket{\beta}$ and $\hat{D}^{\dagger}\left(\beta\right) \hat{a}^{\dagger} \hat{D}\left(\beta\right)=\hat{a}^{\dagger} + \beta^{*}$. For simplicity, we have taken $\alpha$ as a real number, but all calculation can be done with a $\alpha$ complex .

\subsection{Second order correction}
Let us now turn to get the second perturbative order solution by using again the general solution \eqref{a7}, but now running $k=2$; we obtain the set of equations
\begin{equation} \label{c19}
\ket{\psi(\tau)}\approx N^{(2)}(\tau)\left[\ket{\psi^{(0)}} + \lambda \ket{\psi^{(1)}} + \lambda^2 \ket{\psi^{(2)}} \right],
\end{equation}
where
\begin{subequations}
	\begin{align} \label{c20a}
	\ket{\psi^{(2)}}&=-e^{-i\hat{H}_{0} \tau } \int_{0}^{\tau} e^{i\hat{H}_{0} \tau_{1} } \left(\hat{n} + \frac{\kappa}{2} \hat{\sigma_{z}}\right) e^{-i\hat{H}_{0} \tau_{1} } \int_{0}^{\tau_{1}}  e^{i\hat{H}_{0} \tau_{2} } \left(\hat{n} + \frac{\kappa}{2} \hat{\sigma_{z}}\right) e^{-i\hat{H}_{0} \tau_{2} } d\tau_{2} d\tau_{1} \ket{\psi(0)} , \\ \label{c20b}
	N^{(2)}(\tau)^{-2}&=1+ \lambda^2 \left[ 2 \Re \left( \braket{\psi^{(1)}}{\psi^{(2)}} \right) + \braket{\psi^{(1)}}{\psi^{(1)}}\right] + 2 \lambda^3 \Re \left(\braket{\psi^{(1)}}{\psi^{(2)}}\right) + \lambda^4 \left(\braket{\psi^{(2)}}{\psi^{(2)}}\right) .
	\end{align}
\end{subequations} 
We insert Eq.~\eqref{c4} into \eqref{c20a} and after integration one gets
\begin{align} \label{c21}
\ket{\psi^{(2)}}=& \frac{e^{-i\hat{H}_{0} \tau }}{4}  \Bigg\lbrace -\left[4 \sin^4 \left( \tau\right) + \sin^2(2\tau)\right] \hat{O}_7 + \left( \hat{O}_5-\hat{O}_2+\hat{O}_8\right) \sin(2\tau) + \left(4\hat{O}_4 + 2 \hat{O}_6-\hat{O}_3 \right) \sin^2(\tau) \nonumber \\
& + \tau \left[-2\hat{O}_8 -2\hat{O}_5+2\cos(2\tau)\hat{O}_2 + \sin(2\tau) \left(\hat{O}_3-2\hat{O}_6\right) \right] - \tau^2 \left(2 \hat{O}_1+\hat{O}_3\right) \Bigg\rbrace \ket{\psi(0)},
\end{align}
with
\begin{subequations}
\begin{align} \label{c22}
\hat{O}_1=& \hat{n}^2 +\frac{\eta}{2}\hat{n}\left[\eta + i \left(\hat{a}-\hat{a}^{\dagger}\right)\hat{\sigma_{z}}\right] 
\\
\hat{O}_2=&i 
\left[\hat{\sigma^{+}} \hat{D}\left(i\eta\right)+\hat{\sigma^{-}} \hat{D}^{\dagger}\left(i\eta\right)\right]
\Bigg\lbrace
 -\eta^2\hat{n}+\frac{\eta^2}{4}\left(\hat{a}^2+\hat{a}^{\dagger2}\right) - \frac{\eta^4}{4} -\frac{\eta^2}{4}
-i\frac{\eta}{2} \left(\hat{a}-\hat{a}^{\dagger}\right)\left(\hat{n} + \eta^2 \right) \hat{\sigma}_{z}
\nonumber\\
&+\left[   \frac{\kappa}{2} \hat{n} 
+  \frac{\eta^2 \kappa}{4} + i \frac{\eta \kappa}{4} \left(\hat{a}-\hat{a}^{\dagger}\right) \hat{\sigma}_{z}\right]  \hat{\sigma}_{z}\Bigg\rbrace 
\\
\hat{O}_3=& \frac{\eta^2}{2}\left[4 \hat{n}-\left(\hat{a}^2+\hat{a}^{\dagger2}\right) + \eta^2 + 1 \right] + i\eta\left(\hat{a}-\hat{a}^{\dagger}\right)\left( \hat{n} + \eta^2 \right)\hat{\sigma_{z}} \\
\hat{O}_4=&\frac{\eta}{4} \hat{n}\left[\eta + i \left(\hat{a}-\hat{a}^{\dagger}\right) \hat{\sigma_{z}}-\frac{\kappa}{\eta} \hat{\sigma_{z}} \right] \\
\hat{O}_5=&\hat{O}_2+  
\left[\hat{\sigma^{+}} \hat{D}\left(i\eta\right)+\hat{\sigma^{-}} \hat{D}^{\dagger}\left(i\eta\right)\right]
\Bigg\lbrace
i \frac{\eta^2}{2} \left[\eta + i \left(\hat{a}-\hat{a}^{\dagger}\right) \hat{\sigma_{z}}-\frac{\kappa}{\eta} \hat{\sigma_{z}} \right]^2
\nonumber \\ &
-i\frac{\eta^2}{4}  \left[\eta + i \left(\hat{a}-\hat{a}^{\dagger}\right) \hat{\sigma_{z}}-\frac{\kappa}{\eta} \hat{\sigma_{z}} \right] \left[\eta + i \left(\hat{a}-\hat{a}^{\dagger}\right) \hat{\sigma_{z}}\right] 
-\frac{\eta}{2}  \left(\hat{a}+\hat{a}^{\dagger}\right) \hat{\sigma_{z}}  \Big\rbrace
\\
\hat{O}_6=&\frac{\kappa}{4} \Bigg\lbrace 2\hat{n} + \eta\left[\eta+i\left(\hat{a}-\hat{a}^{\dagger}\right) \hat{\sigma_{z}}\right] \Bigg\rbrace  \hat{\sigma_{z}} \\
\hat{O}_7=& \frac{\kappa}{8}\Big\lbrace \kappa \hat{\sigma_{z}}-\eta\left[\eta + i \left(\hat{a}-\hat{a}^{\dagger}\right) \hat{\sigma_{z}}\right] \Big\rbrace \hat{\sigma_{z}} \\
\hat{O}_8=&-2i\left[\hat{\sigma^{+}} \hat{D}\left(i\eta\right)+\hat{\sigma^{-}} \hat{D}^{\dagger}\left(i\eta\right)\right] \hat{O}_7
\end{align}
\end{subequations} 
Applying the unperturbed evolution operator $e^{-i\hat{H}_{0}\tau}$ and after some algebraic manipulation, we arrive to
\begin{align} \label{c23}
\ket{\psi^{(2)}}=& -\frac{\tau}{8} \cos(\tau) \Big\lbrace 4 \tau \hat{n}^2 + \eta^2 \left[\tau-\tan(\tau)\right] \left(6 \hat{n} + \eta^2 +1\right)  - \eta^2\left[ \tau-\tan(\tau)\right] \left(\hat{a}^2 + \hat{a}^{\dagger2}\right) \Big\rbrace \ket{\psi(0)} \nonumber\\
&-i\frac{\eta^2}{8} \cos(\tau) \left[\tau-\tan(\tau)\left(1+\tau^2\right)\right] \left[\hat{\sigma^{+}} \hat{D}\left(i\eta\right)+\hat{\sigma^{-}} \hat{D}^{\dagger}\left(i\eta\right)\right] \left(2\hat{n} + \eta^2 +1\right)\ket{\psi(0)} \nonumber\\ 
& + i \frac{\eta^2}{8} \cos(\tau)  \left[\tau-\tan(\tau)\left(1+\tau^2\right)\right] \left[\hat{\sigma^{+}} \hat{D}\left(i\eta\right)+\hat{\sigma^{-}} \hat{D}^{\dagger}\left(i\eta\right)\right] \left(\hat{a}^2 + \hat{a}^{\dagger2}\right)  \ket{\psi(0)} \nonumber\\
& -i\frac{\kappa \eta}{4} \cos(\tau) \left[\tau- \tan(\tau)\right] \left[\hat{\sigma^{+}} \hat{D}\left(i\eta\right)-\hat{\sigma^{-}} \hat{D}^{\dagger}\left(i\eta\right)\right] \left[\eta + i \left(\hat{a}-\hat{a}^{\dagger}\right) \hat{\sigma}_{z}\right] \ket{\psi(0)}  \nonumber\\
& -\frac{\eta^3}{4}  \cos(\tau) \left[\tau-\tan(\tau)\left(1+\tau^2\right)\right] \left[\hat{\sigma^{+}} \hat{D}\left(i\eta\right)-\hat{\sigma^{-}} \hat{D}^{\dagger}\left(i\eta\right)\right] \left(\hat{a}-\hat{a}^{\dagger}\right) \ket{\psi(0)} \nonumber\\
&-\frac{\eta}{4}\cos(\tau) \left[\tau-\tan(\tau)\left(1-\tau^2\right)\right] \left[\hat{\sigma^{+}} \hat{D}\left(i\eta\right)-\hat{\sigma^{-}} \hat{D}^{\dagger}\left(i\eta\right)\right] \left(\hat{a} + \hat{a}^{\dagger}\right)  \ket{\psi(0)} \nonumber\\
&-i\frac{\eta \tau }{4} \cos(\tau) \left[ \tau-\tan(\tau)\right] \left[ \hat{a}\left(2\hat{n}+\eta^2-1\right)-\hat{a}^{\dagger}\left(2\hat{n}+\eta^2+1\right)\right] \hat{\sigma_{z}} \ket{\psi(0)} \nonumber\\
& -i\frac{\kappa^2}{8} \cos(\tau) \left[\tau- \tan(\tau)\right] \left[\hat{\sigma^{+}} \hat{D}\left(i\eta\right)+\hat{\sigma^{-}} \hat{D}^{\dagger}\left(i\eta\right)\right] \ket{\psi(0)}  \nonumber\\
& + \frac{\eta \tau^2}{2} \sin(\tau) \left[\hat{\sigma^{+}} \hat{D}\left(i\eta\right)-\hat{\sigma^{-}} \hat{D}^{\dagger}\left(i\eta\right)\right] \left(\hat{a}-\hat{a}^{\dagger}\right) \hat{n} \ket{\psi(0)} \nonumber\\
& + i \frac{\tau^2}{2} \sin(\tau) \left[\hat{\sigma^{+}} \hat{D}\left(i\eta\right)+\hat{\sigma^{-}} \hat{D}^{\dagger}\left(i\eta\right)\right] \hat{n} \left(\hat{n} + \eta^2\right)\ket{\psi(0)} \nonumber\\
&-\frac{\kappa}{8} \left(4 \hat{n} \hat{\sigma}_{z} +\kappa\right) \tau \sin(\tau) \ket{\psi(0)}.
\end{align}
Once the second order term has been calculated, the normalization constant and the complete solution can be obtained using Eq.~\eqref{c20a} and Eq.~\eqref{c20b}. For practical purpose, let us consider the initial condition $\ket{\psi(0)}=\ket{i\alpha}\ket{e}$, then, the solution at second order is given by
\begin{align}\label{c24}
\ket{\psi(\tau)}_{i\alpha, e}=& N^{(2)}_{i\alpha,e} \cos(\tau) \bigg( 1+  i\frac{\lambda \eta}{2} \bigg\lbrace \left(\alpha-\eta\right)  \left[\tau-\tan(\tau)\right]  - \frac{\kappa}{\eta} \tan(\tau)  \bigg\rbrace \bigg) \ket{i\alpha} \ket{e}-\lambda^2 N^{(2)}_{i\alpha,e} F_{1}\left(\alpha,\tau \right) \ket{i\alpha} \ket{e}\nonumber\\
& -i\frac{\lambda}{2} N^{(2)}_{i\alpha,e} (\tau)  \bigg\lbrace \cos(\tau) \Big[\left(2\alpha-\eta\right)\tau + \eta \tan(\tau)\Big]-2i\lambda F_{2}\left(\alpha,\tau \right) \bigg\rbrace \left(\frac{\partial}{\partial \alpha} + \alpha\right) \ket{i\alpha}\ket{e} \nonumber\\
& -\frac{\lambda}{2} N^{(2)}_{i\alpha,e}(\tau) \Big[ \left(2\alpha-\eta\right) \tau \sin(\tau) +2i\lambda F_{5}\left(\alpha,\tau \right) \Big]  \left[\frac{\partial}{\partial \left(\alpha-\eta\right)} + \alpha-\eta\right] \ket{i\left(\alpha-\eta\right)} \ket{g} \nonumber\\
&-i\lambda^2 N^{(2)}_{i\alpha,e} (\tau) F_{6}\left(\alpha,\tau \right) \left[\frac{\partial^2}{\partial^2 \left(\alpha-\eta\right)} + 2 \left(\alpha-\eta\right) \frac{\partial}{\partial \left(\alpha-\eta\right)} + \left(\alpha-\eta\right)^2\right] \ket{i\left(\alpha-\eta\right)} \ket{g}\nonumber\\
& - i N^{(2)}_{i\alpha,e} (\tau) \left[\sin(\tau)\left(1-i \frac{\alpha \eta \lambda \tau}{2}\right)-\lambda^2 F_{4}\left(\alpha,\tau \right)\right]  \ket{i\left(\alpha-\eta\right)} \ket{g} \nonumber\\
& -\lambda^2 N^{(2)}_{i\alpha,e} (\tau) F_{3}\left(\alpha,\tau \right) \left[\frac{\partial^2}{\partial^2 \alpha} + 2\alpha \frac{\partial}{\partial \alpha} + \alpha^2\right] \ket{i\alpha} \ket{e},
\end{align}
where
\begin{align} \label{c25}
\left[ N^{(2)}_{i\alpha,e}\right]^{-2}= & 1 -\frac{\tau^2 \lambda^2}{4} \alpha \eta \left(\eta^2 +1\right) \sin^2\left(\tau\right) + \lambda^4 \Big\lbrace  F^2_{1}\left(\alpha,\tau \right) +  F^2_{2}\left(\alpha,\tau \right) + 2\left[ F^2_{3}\left(\alpha,\tau \right) +  F^2_{6}\left(\alpha,\tau \right) \right] +  F^2_{4}\left(\alpha,\tau \right) +  F^2_{5}\left(\alpha,\tau \right)\Big\rbrace \nonumber\\
& + \alpha^2 \lambda^4 \Big\lbrace 2\left[ F_{1}\left(\alpha,\tau \right) + 2 F_{3}\left(\alpha,\tau \right)\right] F_{3}\left(\alpha,\tau \right) + F^2_{2}\left(\alpha,\tau \right) \Big\rbrace + \alpha^3 \lambda^4 \left[2 F_{2}\left(\alpha,\tau \right) + \alpha F_{3}\left(\alpha,\tau \right)\right] F_{3}\left(\alpha,\tau \right) \nonumber\\
& + 2\alpha \lambda^4 \left[ F_{1}\left(\alpha,\tau \right) + 2 F_{3}\left(\alpha,\tau \right)\right]  F_{2}\left(\alpha,\tau \right) + 2\left(\alpha-\eta\right) \lambda^4 \left[ 2 F_{6}\left(\alpha,\tau \right) - F_{4}\left(\alpha,\tau \right)\right]  F_{5}\left(\alpha,\tau \right) \nonumber\\
& + \left(\alpha-\eta\right)^2 \lambda^4 \Big\lbrace  F^2_{5}\left(\alpha,\tau \right) -2\left[ F_{4}\left(\alpha,\tau \right) - 2 F_{6}\left(\alpha,\tau \right)\right] F_{6}\left(\alpha,\tau \right)  \Big\rbrace \nonumber\\
& + \left(\alpha-\eta\right)^3 \lambda^4 \left[ 2 F_{5}\left(\alpha,\tau \right) + \left(\alpha-\eta\right) F_{6}\left(\alpha,\tau \right)\right] F_{6}\left(\alpha,\tau \right)
\end{align}
with
\begin{subequations}
\begin{align} \label{c26}
F_{1}(\alpha,\tau)=& \frac{\tau \eta}{8} \cos\left(\tau\right) \Bigg\lbrace  \left[\tau-\tan\left(\tau\right)\right] \left[\alpha^2 \eta + \left(\eta^2 + 1\right) \left(\eta - 2 \alpha\right)\right] + \frac{\kappa^2}{\eta}\tan\left(\tau\right) \Bigg\rbrace\\
F_{2}(\alpha,\tau)=& \frac{\tau}{4} \cos\left(\tau\right) \Big\lbrace 2\alpha \tau  + \eta \left[3 \eta \alpha-2\alpha^2-\left(\eta^2+1\right) \right] \left[\tau-\tan\left(\tau\right)\right] + 2 \alpha \kappa \tan\left(\tau\right)  \Big\rbrace \\
F_{3}(\alpha,\tau)=& \frac{\tau}{8} \cos\left(\tau\right) \Bigg\lbrace 4 \tau \alpha^2 + \eta \left(\eta-4\alpha\right) \Big[ \tau  -  \tan\left(\tau\right)  \Big] \Bigg\rbrace \\
F_{4}(\alpha,\tau)=& \frac{\eta}{8} \cos\left(\tau\right) \Bigg\lbrace \tau^2  \left[\alpha \eta \left(\alpha + \eta \right) +3\alpha-\eta \right] \tan\left(\tau\right) -  \Big[ 2 \left(\alpha \kappa + \eta -\alpha\right) + \eta \left(\alpha^2 + \frac{\kappa^2}{\eta^2} +1\right)   \Big] \left[\tau-\tan\left(\tau\right)\right]\Bigg\rbrace \\
F_{5}(\alpha,\tau)=& \frac{\cos\left(\tau\right)}{4} \Big\lbrace  \eta \left(\alpha \eta + \kappa +1\right) \left[\tau-\tan\left(\tau\right)\right] -\left(2\alpha-\eta\right)\left(\alpha \eta  +1\right) \tau^2 \tan\left(\tau\right)\Big\rbrace \\
F_{6}(\alpha,\tau)=& \frac{\cos\left(\tau\right)}{8} \Big\lbrace \tau^2 \left[4 \alpha \left(\eta-\alpha\right) - \eta^2 \right] \tan\left(\tau\right) + \eta^2 \left[\tau-\tan\left(\tau\right)\right]  \Big\rbrace.
\end{align}
\end{subequations}

\subsection{Comparison of the perturbative solution with the small rotation approximation solution}
\begin{figure}[H]
	{\includegraphics[width=0.46\textwidth]{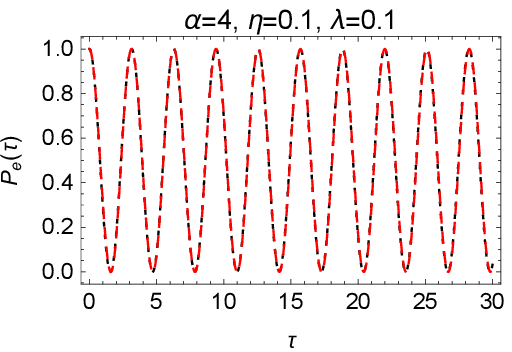}}
	\hfill {\includegraphics[width=0.46\textwidth]{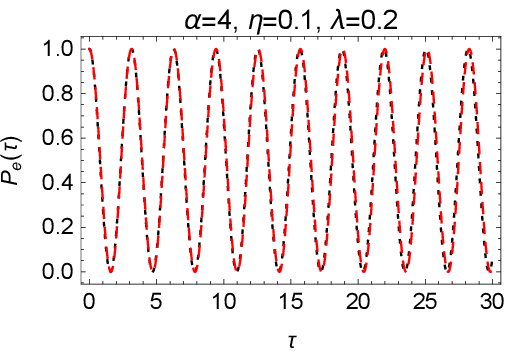}}\hfill {\includegraphics[width=0.46\textwidth]{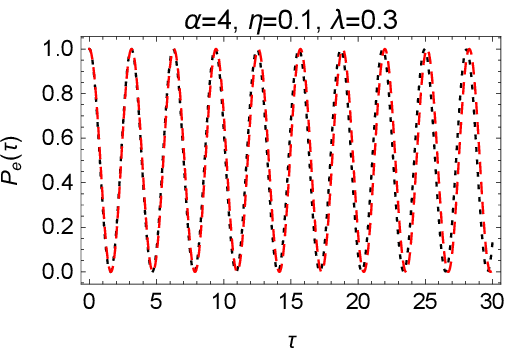}} \hfill {\includegraphics[width=0.46\textwidth]{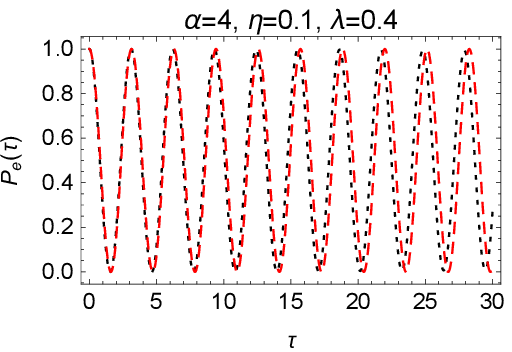}}
	\caption{Plot of $P_{e}(\tau)$ as a function of $\tau$ using $\lambda=0.1,\, 0.2,\, 0.3$ and $ 0.4$ and considering $\kappa=0$, $\eta=0.1$ and $\alpha=4$. The black dotted line represents the small rotation approximation solution and the red dashed line denotes the perturbative solution.}
	\label{fig1}
\end{figure}

In order to verify the validity and accuracy of our perturbative solution, we calculate the probability to find the ion in its excited state, $P_{e}(\tau)=\bra{\psi(\tau)}\ket{e}\bra{e}\ket{\psi(\tau)}$, and compared it with the expression
\begin{equation}  \label{c27}
P_{e}(\tau)_{\textrm{exact}}= \frac{1}{2} \Bigg\lbrace 1+ \exp\left[-2\left(\alpha-\eta/2\right)^2 \sin^2\left(\tau \chi^{\textrm{high}}\right)\right] \cos\left[ \tau \left(2-\chi^{\textrm{high}}\right) -\left(\alpha-\eta/2\right)^2 \sin\left(2\tau \chi^{\textrm{high}}\right)\right] \Bigg\rbrace,
\end{equation}
taken from Eq.16 and Eq.18 of reference\cite{A34}, which is the small rotation approximation solution for this system and where $\chi^{\textrm{high}}=-\lambda^2 \eta^2/2$ in the case of high intensity regime. Hence, using Eq.~\eqref{c24} to calculate $P_{e}(\tau)$ yields the following expression
\begin{align}  \label{c28}
P_{e}(\tau)=&\left[ N^{(2)}_{i\alpha,e}\right]^{-2} \Bigg\lbrace 
-\lambda^2 \cos\left(\tau\right) \left[ 2 \alpha^2 F_{3}\left(\alpha,\tau\right) + \alpha F_{2} \left(\alpha,\tau\right) + 2 F_{1} \left(\alpha,\tau\right)\right] + \lambda^4 \left[ F_{1}^2 \left(\alpha,\tau\right) +  F_{2}^2 \left(\alpha,\tau\right) + 2 F_{3}^2 \left(\alpha,\tau\right) \right]  \nonumber \\
& + \cos^2\left(\tau\right) + \lambda^2 \left[ g_{1}^2\left(\alpha,\tau\right) + \left(\alpha^2 + 1\right) g_{2}^2\left(\alpha,\tau\right) -2 \alpha g_{1}\left(\alpha,\tau\right) g_{2}\left(\alpha,\tau\right) \right] + 2 \lambda^4 \alpha^3  F_{2} \left(\alpha,\tau\right)  F_{3} \left(\alpha,\tau\right) \Bigg\rbrace \nonumber \\
& + \alpha^2 \lambda^4 \left[ N^{(2)}_{i\alpha,e}\right]^{-2} \Bigg\lbrace  F_{2}^2 \left(\alpha,\tau\right) + 2 F_{3} \left(\alpha,\tau\right) \left[  F_{1} \left(\alpha,\tau\right) + 2 F_{3} \left(\alpha,\tau\right) \right] \Bigg\rbrace \nonumber \\
& +  \alpha \lambda^4 \left[ N^{(2)}_{i\alpha,e}\right]^{-2} \Big\lbrace 2 F_{2} \left(\alpha,\tau\right) \left[ F_{1} \left(\alpha,\tau\right) + 2 F_{3} \left(\alpha,\tau\right)\right]  + \alpha^3 F_{3}^2 \left(\alpha,\tau\right) \Big\rbrace
\end{align}
where
\begin{subequations}
\begin{align} \label{c29}
g_{1} \left(\alpha,\tau\right)=& \frac{\eta \cos\left(\tau\right)}{2} \Bigg\lbrace  \left( \alpha-\eta\right) \left[\tau-\tan\left(\tau\right)\right] -\frac{\kappa}{\eta} \tan\left(\tau\right) \Bigg\rbrace\\
g_{2}\left(\alpha,\tau\right)=& \frac{\cos\left(\tau\right)}{2} \Big\lbrace 2\alpha \tau -\eta \left[ \tau- \tan\left(\tau\right)\right] \Big\rbrace.
\end{align}
\end{subequations}
The $P_{e}(\tau)$ obtained by our approach, Eq.~\eqref{c28}, and those from the small rotation approximation solution are plotted in Fig.~\ref{fig1} for several values of the perturbative parameter $\lambda$.  It is clear that the perturbative results, indicated by the red dashed line, are sufficiently accurate to reproduce the small rotation approximation solution denoted by the black dotted line, provided condition $\lambda \tau \ll 1$ holds. Otherwise, it is logical to expect that when $\lambda \tau \gg 1$ a substantial difference will arise between the perturbative solution and the small rotation approximation solution.

\section{Perturbative solution for the Rabi model}
As we commented at the end of Section 3, it has been shown in \cite{A7.2} that the ion-trap system is formally equivalent to the quantum Rabi model when we consider the unitary transformation
\begin{equation}
\hat{T}=\frac{1}{2\sqrt{2}}\left[\hat{D}^{\dagger}\left(i\eta/2\right) + \hat{D}\left(i\eta/2\right)\right] \hat{I} + \frac{1}{2\sqrt{2}}\left[\hat{D}^{\dagger}\left(i\eta/2\right) - \hat{D}\left(i\eta/2\right)\right]\hat{\sigma}_{z} + \frac{1}{\sqrt{2}}\left[\hat{\sigma}^{+} \hat{D}\left(i\eta/2\right)-\hat{\sigma}^{-} \hat{D}^{\dagger}\left(i\eta/2\right)\right],  
\end{equation}
 and we take $ \nu\rightarrow \omega$, $\Omega \rightarrow \frac{\omega_{0}}{2}$ and $\frac{\eta \nu}{2} \rightarrow g $; when this transformation is done, we get the transformed Schrödinger equation
\begin{equation} \label{b3}
i\frac{d}{dt}\ket{\phi(t)}=\hat{\mathcal{H}}_{\textrm{Rabi}} \ket{\phi(t)},
\end{equation}
where
\begin{equation}
    \hat{\mathcal{H}}_{\textrm{Rabi}}=\hat{T} \hat{H}_{\textrm{ion}} \hat{T}^\dagger=\omega \hat{n}+ \frac{\omega_{0}}{2} \hat{\sigma_{z}} +i g\left( \hat{a}-\hat{a}^\dagger\right) \left( \hat{\sigma}^{+} + \hat{\sigma}^{-} \right).
\end{equation}
Hence, we can perform the transformation $\ket{\phi(t)}=\hat{T}\ket{\psi(t)}$ and also get perturbative solutions of the quantum Rabi model for the weak $\left(g/\omega\right)\ll 1$ and strong $\left(g/\omega\right)\gg1$ coupling regime. As the low intensity regime case has been considered extensively  \cite{1A,2A,3A,4A,5A,6A,A7.1,A7.2}, we focus now on the high intensity regime perturbative solution. Taking advantage of this equivalence and applying $\hat{T}$ to expression \eqref{c24}, one gets
\begin{align}
\ket{\phi(t)} & \approx - N^{(2)}_{i\alpha,e} \Bigg\lbrace \cos(\tau) + i\lambda \left[g_{1}\left(\alpha,\tau\right)-\frac{\eta}{2} g_{2}\left(\alpha,\tau\right)\right]-\lambda^2 \left[ F_{1}\left(\alpha,\tau\right) + \frac{\eta}{2} F_{2}\left(\alpha,\tau\right) + \frac{\eta^2}{4} F_{3}\left(\alpha,\tau\right)\right] \Bigg\rbrace \ket{-} \ket{\gamma}
\nonumber\\
& + N^{(2)}_{i\alpha,e} \Bigg\lbrace i \lambda^2 \left[ F_{4}\left(\alpha,\tau\right) + \frac{\eta}{2}  F_{5}\left(\alpha,\tau\right) -\frac{\eta^2}{4}  F_{6}\left(\alpha,\tau\right) \right]-i\sin\left(\tau\right) - \lambda \left[  g_{3}\left(\alpha,\tau\right)-\frac{\eta}{2}  g_{4}\left(\alpha,\tau\right) \right]\Bigg\rbrace  \ket{+} \ket{\gamma} \nonumber\\
& + N^{(2)}_{i\alpha,e} \Bigg\lbrace \lambda^2 \left[ F_{2}\left(\alpha,\tau\right) + \eta F_{3}\left(\alpha,\tau\right) \right]  + i \lambda  g_{2}\left(\alpha,\tau\right) \Bigg\rbrace \left(\frac{\partial}{\partial \gamma} + \gamma \right)\ket{-} \ket{\gamma} \nonumber\\
& - N^{(2)}_{i\alpha,e} \Bigg\lbrace \lambda  g_{4}\left(\alpha,\tau\right) + i \lambda^2 \left[F_{5}\left(\alpha,\tau\right) - \eta F_{6}\left(\alpha,\tau\right) \right]   \Bigg\rbrace \left(\frac{\partial}{\partial \gamma} + \gamma \right)\ket{+} \ket{\gamma} \nonumber\\
& + \lambda^2 N^{(2)}_{i\alpha,e} \left[ F_{3}\left(\alpha,\tau\right) \ket{-}-i F_{6}\left(\alpha,\tau\right) \ket{+} \right] \left(\frac{\partial^2}{\partial^2 \gamma} +2\gamma \frac{\partial}{\partial\gamma} + \gamma^2 \right) \ket{\gamma}
\end{align}
which is the second order perturbative solution for the Rabi model and where $\ket{\pm}=\frac{1}{\sqrt{2}}\left[\ket{g} \pm \ket{e}\right]$,  $\gamma=i\left(\alpha-\eta/2\right)$,  $g_{3}\left(\alpha,\tau\right)=\frac{\tau \alpha \eta}{2} \sin\left(\tau\right)$ and $g_{4}\left(\alpha,\tau\right)=\frac{\tau}{2}\left( 2\alpha-\eta\right)\sin\left(\tau\right)$.

\section{Conclusions}
In contrast to standard perturbative approaches with rather cumbersome algebra and based on special assumptions to reach an approximate solution, we conclude that the NMPM enables us to work out a simple  perturbatively treatment for a trapped-ion system interacting with a laser field in the high intensity regime. The perturbative solution has been shown to be capable to reproduce, with high accuracy and self-consistently, the results from the small rotation approximation solution reported in the literature. Indeed, our work could pave the way to study the perturbative solutions for an $N$-ion system instead a single ion.

\section{Acknowledgment}
B.M. Villegas-Martínez wish to express his gratitude to CONACyT as well as to the National Institute of Astrophysics, Optics and Electronics INAOE for financial support.


\begin{thebibliography}{unsrt}

\bibitem{1A} D.M. Meekhof, C. Monroe, B.E. King, W.M. Itano and D.J. Wineland, Phys.Rev.Lett. 761796 (1996).
\bibitem{2A} S. Wallentowitz and W. Vogel, Phys. Rev. A 55 4438 (1997).
\bibitem{3A} S. Wallentowitz, W. Vogel and P.L. Knight Phys.Rev. A 59 531 (1999).
\bibitem{4A} Z. Kis, W. Vogel and L. Davidovich, Phys. Rev. A 64 033401 (2001).
\bibitem{5A} R.L. de Matos Filho and W. Vogel, Phys.Rev.Lett. 76 608 (1996).
\bibitem{6A} R.L. de Matos Filho and W. Vogel, Phys. Rev.A 54 4560 (1996).
\bibitem{A7.1} H. Moya-Cessa and P. Tombesi,  Phys. Rev. A 61, 025401 (2000).
\bibitem{A7.2} J. Casanova, R. Puebla, H. Moya-Cessa and M.B. Plenio, npj Quantum Information 4, 47 (2018).
\bibitem{A7} D. Leibfried, D. Meekhof, B.E. King, C. Monroe, W.M. Itano and D.J. Wineland, Phys.Rev. Lett. 77 4281 (1996).
\bibitem{A8} C. Ospelkaus, et al, Phys. Rev. Lett. 101, 090502, 1–4 (2008).
\bibitem{A9} E. Solano, R.L. de Matos Filho, and N. Zagury, Phys. Rev. A59, R2539–R2543 (1999).
\bibitem{A10} A. Barenco, et al, Phys. Rev. A52, 3457–3467 (1995).	
\bibitem{A11} D. Jonathan D, M.B. Plenio and P.L. Knight, Phys.Rev. A 62, 042307 (2000).
\bibitem{A12} L. Allen, and J.H. Eberly, \textit{Optical Resonance and two-level Atoms}, chap.2. Wiley, New York (1975).
\bibitem{A13} J.I. Cirac, R. Blatt, A.S. Parkins, and P. Zoller, Phys. Rev. Lett. 70, 762 (1993).
\bibitem{A14} J.I. Cirac, R. Blatt, A.S. Parkins, and P. Zoller, Phys. Rev. A49, 1202 (1994).
\bibitem{A15} L.M. Duan.Phys. Rev. Lett., 93:100502 (2004).
\bibitem{A16} R.L. Taylor, C.D. B. Bentley, J.S. Pedernales, L. Lamata, E.Solano, A.R.R. Carvalho, and J.J. Hope,arXiv:1601.00359 (2016).
\bibitem{A17} R. Puebla, M.J. Hwang, J. Casanova, and M. B. Plenio, Phys. Rev. A95, 063844 (2017).
\bibitem{A18} P. Aniello et al., quant-ph/0301138 (2003).
\bibitem{A19} P. Aniello, A. Porzio and S. Solimeno, J. Opt. B: Quan-tum Semiclass. Opt.5, S233 (2003)
\bibitem{A20} P. Aniello, J. Opt. B Quantum Semiclass. Opt., V.7, S507–S522, quant-ph/0508017 (2005).
\bibitem{A21} J. Martínez-Carranza, F. Soto-Eguibar, H. Moya-Cessa, Eur. Phys. J. D. 66(1), 1-6, 71 (2012).
\bibitem{A22} J. Martínez-Carranza, H. Moya-Cessa, F. Soto-Eguibar, \textit{La teoría de perturbaciones en la mecánica cuántica}, Editorial Académica Española, (2012).
\bibitem{A23} B. M.  Villegas-Martínez, F.  Soto-Eguibar, H.M.  Moya-Cessa, Adv.  Math.  Phys. 9265039 (2016)
\bibitem{A24} B. M.  Villegas-Martinez, H. M. Moya-Cessa, F. Soto-Eguibar, Journal of Modern Optics. (2018).
\bibitem{A25} B.M. Villegas-Martínez, H.M. Moya-Cessa, F. Soto-Eguibar, Eur. Phys. J. D 74, 137 (2020). 
\bibitem{A26} M. Frasca, Phys. Rev. A58, 3439 (1998).
\bibitem{A27} M. Frasca, Proc. R. Soc. A, 463, 2195-2200 (2007).
\bibitem{A28} M. Frasca, Int. J. Mod. Phys. D15, 1373 (2006).
\bibitem{A29} M. Frasca, Phys. Rev. D73, 027701 (2006).
\bibitem{A30} M. Frasca, Phys. Rev. A43, 45 (1992).
\bibitem{A31} M. Frasca, Phys. Rev. A47, 2374 (1993).
\bibitem{A32} J.F. Poyatos, J.I. Cirac, R. Blatt, and P. Zoller, Phys. Rev. A,54, 1532 (1996).
\bibitem{A33} R. Puri, Mathematical Methods of Quantum Optics, Springer Verlag, (2011).
\bibitem{A34} A. Zuñiga-Segundo, R. Juarez-Amaro, J. M.Vargas-Martinez and H. Moya-Cessa, Ann. Phys,524, 107–111 (2012).

\end{thebibliography}
\end{document}